\documentstyle[stwol,epsf]{article}

\newcommand{\AmS}{{\protect\the\textfont2
  A\kern-.1667em\lower.5ex\hbox{M}\kern-.125emS}}
\newcommand{\figI}      {mttb}                % m_top vs. tanb
\newcommand{\figIII}    {chi2_gyebdl}              % countours
        
\newcommand{\bq}{\begin{equation}}
\newcommand{\eq}{\end{equation}}
\newcommand{\beq}  {\begin{eqnarray}}
\newcommand{\eeq}  {\end{eqnarray}}
\newcommand{\rG}   {{\rm GUT}}
\newcommand{\MG}   {{\ifmmode M_\rG         \else $M_\rG$          \fi}}
\newcommand{\mb}   {{\ifmmode m_{b}         \else $m_{b}$          \fi}}
\newcommand{\mt}   {{\ifmmode m_{t}         \else $m_{t}$          \fi}}
\newcommand{\agut} {{\ifmmode \alpha_\rG    \else $\alpha_\rG$     \fi}}
\newcommand{\mgut} {{\ifmmode M_\rG         \else $M_\rG$          \fi}}
\newcommand{\mze}  {{\ifmmode m_0           \else $m_0$            \fi}}
\newcommand{\mha}  {{\ifmmode m_{1/2}       \else $m_{1/2}$        \fi}}
\newcommand{\tb}   {{\ifmmode \tan\beta     \else $\tan\beta$      \fi}}
\newcommand{\mz}   {{\ifmmode M_{Z}         \else $M_{Z}$          \fi}}
\newcommand{\ai}   {{\ifmmode \alpha_i      \else $\alpha_i$       \fi}}
\newcommand{\aii}  {{\ifmmode \alpha_i^{-1} \else $\alpha_i^{-1}$  \fi}}
\newcommand{\MSb}  {{\ifmmode \overline{\rm MS} \else
                             $\overline{\rm MS}$                   \fi}}
\newcommand{\DRb}  {{\ifmmode \overline{\rm DR} \else
      $\overline{\rm DR}$                   \fi}}
\newcommand{\DRbar}{{\ifmmode \overline{DR} \else $ \overline{DR}$ \fi}}
\newcommand{\msusy}{{\ifmmode M_{SUSY}      \else $M_{SUSY}$       \fi}}
\newcommand{\as}   {{\ifmmode \alpha_s      \else $\alpha_s$       \fi}}
\newcommand{\asmz} {{\ifmmode \alpha_s(M_Z) \else $\alpha_s(M_Z)$  \fi}}
\newcommand{\tal}  {{\ifmmode \tilde{\alpha} \else $\tilde{\alpha}$ \fi}}
\newcommand{\rb}[1]{\raisebox{1.5ex}[-1.5ex]{#1}}

\newcommand{\sws}  {{\ifmmode \;\sin^2\theta_W
                     \else    $\;\sin^{2}\theta_{W}$               \fi}}
\newcommand{\cws}  {{\ifmmode \;\cos^2\theta_W  
                     \else    $\;\cos^{2}\theta_{W}$               \fi}}
\newcommand{\sw}   {{\ifmmode\;\sin\theta_W\else $\sin\theta_{W}$  \fi}}
\newcommand{\cw}   {{\ifmmode\;\cos\theta_W\else $\;\cos\theta_{W}$\fi}}
\newcommand{\tw}   {{\ifmmode\;\tan\theta_W\else $\;\tan\theta_{W}$\fi}}
\newcommand{\bsg}  {{\ifmmode b\rightarrow s\gamma
                     \else $b\rightarrow s\gamma$ \fi}}
\newcommand{\Bbsg}  {{\ifmmode BR(\b\rightarrow s\gamma)
\else $BR(b\rightarrow s\gamma)$ \fi}}

\newcommand{\rPL}  {{\rm Planck}}
\newcommand{\mplanck} {{\ifmmode M_\rPL         \else $M_\rPL$          \fi}}
\newcommand{\rST}  {{\rm SO(10)}}
\newcommand{\msoten} {{\ifmmode M_\rST         \else $M_\rST$          \fi}}

\def\be{\begin{equation}}
\def\ee{\end{equation}}
\def\bea{\begin{eqnarray}}
\def\eea{\end{eqnarray}}
\begin{document}
%hep-ph/yyddnnn  
\title{
%\begin{flushright}
%\vspace*{-2.6cm}
%\noindent
% \hfill IEKP-KA/96-10   \\
% \hfill October 15th, 1996 \\
%\end{flushright}
The Constrained MSSM Revisited}

\author{W. DE BOER }

\address{Inst. f\"ur Experimentelle Kernphysik, Univ. of Karlsruhe, \\ 
        Postfach 6980, 76128 Karlsruhe, Germany}


\twocolumn[\maketitle
\abstracts{Within the Constrained Minimal Supersymmetric Model (CMSSM) it is
possible to predict the low energy gauge couplings and masses of the
3~generation particles from a few supergravity inspired parameters at the GUT scale.
Moreover, the CMSSM predicts electroweak symmetry breaking due to
large radiative corrections from the Yukawa couplings, thus relating
the $Z^0$ boson mass to the top quark mass via the renormalization
group equations (RGE). % \cite{bek1}.  
In addition, the cosmological constraints on the lifetime of the universe
are considered in the fits.  The new precise measurements of
 the strong coupling constant and the top mass as well as  higher
 order calculations of the $\bsg$ rate  exclude perfect fits in the CMSSM,
although  the discrepancies from the best fit parameters are below the $2\sigma$
 level.
}]
\section{The Constrained MSSM }
The Minimal Supersymmetric Standard Model (MSSM) \cite{rev} has become
the leading candidate for a low energy theory consistent with the GUT
requirements. At this conference several new results have been presented,
 which are crucial for the consistency checks of GUT's.
First of all, the $\alpha_s$ crisis has disappeared, since the LEP value went
down  and the DIS measurement as well as the value froms lattice
 calculations went up, 
and the  error on the strong coupling constant has come down  to
 an astonishing low level of about 3\%\cite{schmelling}.
In this analysis I will use for the coupling constants
$\as=0.120\pm0.003$ and $\sin^2\Theta_{\overline{MS}}=0.2317\pm0.004$, which
are the global fit values\cite{hollik1} including the    
 top mass from the combined data of CDF and D0,
 ($175\pm6$ GeV\cite{blondel}) and the 
new higher order calculations for the important $b\rightarrow s\gamma$ rate\cite{misiak}.
 The latter indicate that next to leading log (QCD) corrections
increase the SM value by about 10\%. This can be simulated 
in the lowest level calculation by choosing a renormalization scale $\mu=0.65m_b$,
which will be done in the following.
 Here we repeat our previous analysis\cite{wezp} with the new 
input values mentioned above. The input data and fitted parameters have been summarized 
in table \ref{t1}.
  \begin{figure}[t]
\vspace{-1cm}
    \begin{center}
    \leavevmode
    \epsfxsize=8cm
    \epsffile{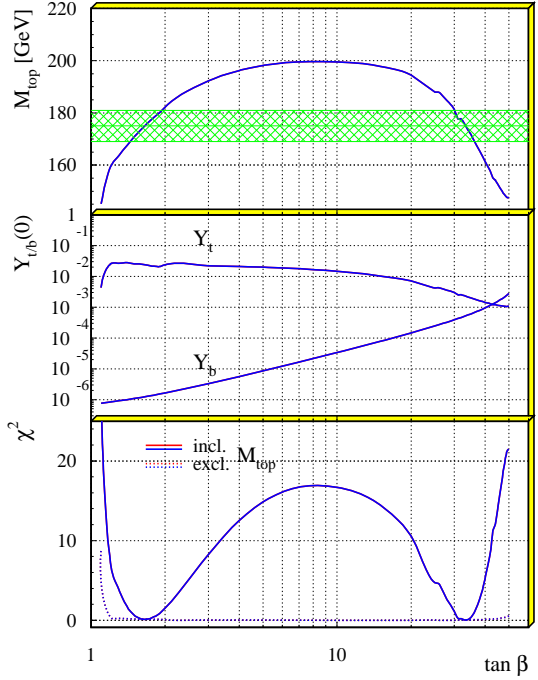}
\end{center}
\vspace{-1.5cm}
  \caption[]{\label{\figI}The top quark mass as function of $\tb$ (top) 
      for values of $\mze,\mha~\approx 1 $ TeV.
      The dependence shown is 
      mainly determined by the $b-\tau $ Yukawa coupling unification.  
      The middle part shows the corresponding values of the Yukawa
      coupling at the GUT scale and the lower part the
      $\chi^2$ values.
      If the top constraint ($\mt=175\pm6$, horizontal band) 
      is not applied, all values of $\tb$ between 1.2 and 50 are allowed
      (thin dotted lines at the bottom), but if the top mass 
      is constrained to the experimental value, only the regions
      $1\le\tb\le3$ and $20\le \tb\le 40$ are allowed.
      }
        %\label{chi2plot}
 \end{figure}
%
%--------------- chi^2
%
  \begin{figure}[t]
    \vspace{-0.51cm}
\begin{center}
    \leavevmode
    \epsfxsize=8.0cm
    \epsffile{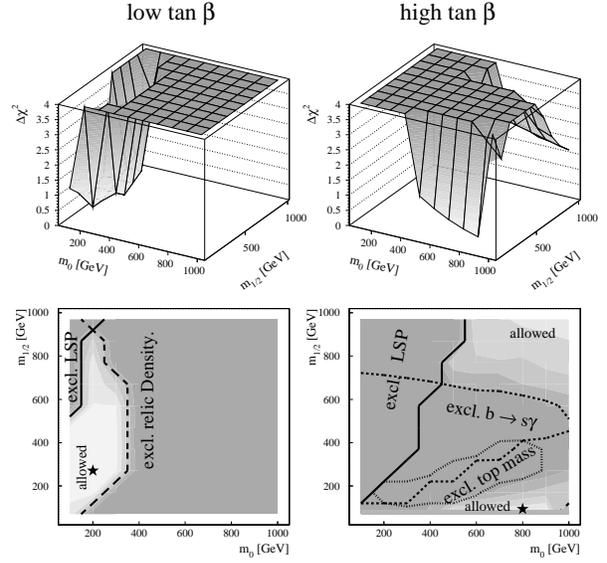}
\end{center}
\vspace{-0.8cm}
\caption[]{\label{\figIII}Contours of the  $\chi^2$-distribution for 
  the low and high $\tb$ solutions.  The different shades indicate
  steps of $\Delta\chi^2 = 1$, so basically only the light shaded
  region is allowed.
      The stars indicate the optimum solution. 
      Contours enclose domains excluded by the particular
      constraints used in the analysis.
}
\end{figure}
%
%--------------- spectrum
%
%\begin{figure}
% \begin{center}
%  \leavevmode
%  \epsfxsize=7cm
%  \epsffile{\figIV.eps}
%\end{center}
%\caption{\label{\figIV}The running of the particle masses and the
%  $\mu$ parameter for low and high $\tb$ values.}
%\end{figure}
{\protect\footnotesize
\begin{table}[t]
\renewcommand{\rb}[1]{\raisebox{1.75ex}[-1.75ex]{#1}}
\begin{center}
\normalsize
\begin{tabular}{|c||c||c|}
\hline
{input data} & {$\Rightarrow$} &Fit parameters \\
\hline
$\alpha_1,\alpha_2,\alpha_3$ &{ min.}   & \mgut,~\agut  \\
\mt   ~\mb,~$m_\tau$                       & { $\chi^2$} & $Y_t^0,~Y_b^0=Y_\tau^0$ \\		
\mz                          &   & $m_0,~m_{1/2},~\mu$,~ \tb     \\
\bsg                         &   & $A_0$  \\
 $\tau_{universe}$           &   &                            \\
\hline
\end{tabular} \end{center}
\caption[]{\label{t1}Summary of  input data and fit parameters, which are
determined from a global fit. They are mainly sensitive to the following input data:
The GUT scale \mgut~ and corresponding coupling constant \agut~
 are  determined from gauge coupling unification,
 the Yukawa couplings $Y_{(t,b,\tau)}^0$ at the GUT scale 
 from the masses of the  3th generation, $\mu$ from electroweak symmetry breaking (EWSB) and
$\tb$ from $b\tau$-unification.
 For 
the low $\tb$ scenario the trilinear coupling  $A_0$ is not very relevant, 
but for large $\tb$ it is determined by $\bsg$ and $b\tau$-unification.
The scalar-  and gaugino  masses $(m_0,m_{1/2})$ enter in all observables.}
\end{table}
}
{\protect\footnotesize
\begin{table}[ht]
\vspace*{0.27cm}
\renewcommand{\arraystretch}{1.30}
\renewcommand{\rb}[1]{\raisebox{1.75ex}[-1.75ex]{#1}}
\begin{center}
\begin{tabular}{|c|r|r|}
\hline
 \multicolumn{3}{|c|}{ Fitted SUSY parameters and masses in GeV}                       \\
\hline
Symbol & \makebox[2.0cm]{\bf{low $\tb$}} & \makebox[2.5cm]{\bf{high $\tb$}}\\
\hline
\vspace{-0.059cm}
 $m_0$,~   $m_{1/2}$        &  230,~ 225            &  850,~115                 \\
\vspace{-0.059cm}
 $\mu(\mz)$,~$\tan\beta$     &  -880,~1.7           & -190,~30   \\
\vspace{-0.059cm}
  $Y_t(\mt)$,~$A_t(\mz)$    & 0.008,~-370           & 0.006,~86             \\
\hline
\vspace{-0.059cm}
  $\tilde{\chi}^0_1$,~$\tilde{\chi}^0_2$         &  96,~194    & 47,~92      \\ 
\vspace{-0.059cm}
  $ \tilde{\chi}^0_3$,~$ \tilde{\chi}^0_4$      &  509,~519   & 414,~417            \\
\vspace{-0.059cm}
  $\tilde{\chi}^{\pm}_1 $,~$\tilde{\chi}^{\pm}_2$ &  194,~518  & 92,~422      \\
\hline
\vspace{-0.059cm}  
  $\tilde{g}$,~$\tilde{q}$,~$\tilde{l}$   &  558,~545,~563  & 300,~885,~854   \\ 
\hline
\vspace{-0.059cm}
  $       h,~H $                   &  74,~673   & 109,~624          \\
\vspace{-0.059cm}
  $       A,~H^\pm $                   &  680,~684  & 624,~630           \\
\hline
\end{tabular} \end{center}
\caption[]{\label{t2}Values of the fitted SUSY parameters (upper part) and
             corresponding susy masses (lower part) 
             for low and high $\tb$ solutions using the new input data discussed
             in the text.   
             }    
\end{table}
}
\section{Results}
The most  restrictive constraints are
the coupling constant unification and the requirement that the
unification scale has to be above $10^{15}$ GeV from the proton
lifetime limits, assuming decay via s-channel exchange of heavy
gauge bosons. They exclude the SM~\cite{abf1} as well
as many other models~\cite{abf2}.
%
%\subsection{Constraints from $b-\tau$ unification}
\vspace{0.2cm}\\
{\bf\underline{Constraints from $b-\tau$ unification}}\vspace{0.2cm}\\
\label{sec:su5}
The requirement of bottom-tau Yukawa coupling unification strongly
restricts the possible solutions in the $\mt$ versus $\tb$ plane.
For $m_t=175\pm6$ GeV only two regions of $\tb$ give an
acceptable $\chi^2$ fit, as shown in the bottom part of
fig.~\ref{\figI}.
The curves in the upper parts are determined by the relations between 
top and bottom masses and $\tb$:
\begin{eqnarray}
m_t^2=4\pi Y_t v^2 \frac{\tan^2\beta}{1+\tan^2\beta}\\
m_b^2=4\pi Y_b v^2 \frac{1}{1+\tan^2\beta}
\end{eqnarray} 
For increasing  $\tb$ $m_t^2$ reaches quickly its plateau $4\pi Y_t v^2$;
 for large $\tb$ $Y_b$ has to compensate the $1/\tan^2\beta$ term,
so it quickly increases (see middle part).
But then the (negative) $Y_b$ contributions to the running of $Y_t$
from loops involving both top and bottom
cannot be neglected anymore, which  decrease
  $Y_t$ and correspondingly the top mass for high $\tb$.
%
%\subsection{Electroweak Symmetry Breaking (EWSB)}
\vspace{0.2cm}\\
{\bf \underline{Electroweak Symmetry Breaking (EWSB)}}\vspace{0.2cm}\\
\label{sec:elwsb}
Radiative corrections can trigger spontaneous symmetry breaking in the
electroweak sector.  In this case the Higgs potential does not have
its minimum for all fields equal zero, but the minimum is obtained for
non-zero vacuum expectation values of the fields.  Solving $\mz$ from
the minimum of the Higgs potential yields: 
\begin{equation}
\label{defmz}
\frac{\mz^2}{2}=\frac{m_1^2+\Sigma _1-(m_2^2+\Sigma _2) \tan^2\beta}
{\tan^2\beta-1},
\end{equation} 
where $m_{1,2}$ are the mass terms in the Higgs potential
and  $\Sigma_1$ and $\Sigma_2$ their radiative corrections.
Note that the radiative corrections are needed, since unification
at the GUT scale with $m_1=m_2$ would lead to $M_Z<0$.
In order to obtain $M_Z>0$ one needs to have  
%
%This equation  can be written as
%\begin{eqnarray} \label{defmz_2}
%  \tan^2\beta & = & \frac{m_1^2+\Sigma_1 +
%    \frac{1}{2}M_Z^2} {m_2^2 + \Sigma_2 + \frac{1}{2}M_Z^2}.
%\end{eqnarray}
%
% At low energy $m_1^2 +\Sigma_1 > m_2^2 + \Sigma_2$,
 which happens at low energy since $\Sigma_2$
($\Sigma_1$) contains large negative corrections proportional to $Y_t$
($Y_b$) and $Y_t \gg Y_b$. 
%Consequently the numerator in
%eq.~\ref{defmz_2} is always larger than the denominator, implying 
%$\tan^2\beta > 1$. 
Electroweak symmetry breaking for the large $\tb$ scenario is not so easy,
since  eq. \ref{defmz} can be rewritten as:
\begin{eqnarray} \label{defmz_2}
  \tan^2\beta & = & \frac{m_1^2+\Sigma_1 +
    \frac{1}{2}M_Z^2} {m_2^2 + \Sigma_2 + \frac{1}{2}M_Z^2}.
\end{eqnarray}
For large $\tb$  $Y_t \approx Y_b$, so $\Sigma_1
\approx \Sigma_2$. Eq.~\ref{defmz_2} then
requires the starting values of $m_1$ and $m_2$ to be different in
order to obtain a large value of $\tan\beta$, which could happen
if the symmetry group above the GUT scale has a larger rank than
the SM, like e.g. SO(10)\cite{pok}. In this case the quartic interaction
 (D-) terms in the Higgs potential can generate quadratic mass terms,
if the Higgs fields develop non-zero v.e.v's after spontaneous symmetry breaking.

Alternatively, one has to assume the simplest GUT group $SU(5)$, which has the
same rank as the SM, so no additional groups are needed to break SU(5)
and consequently no D-terms are generated. In this case EWSB can only be 
generated, if $Y_b$ is sufficiently below $Y_t$, in which case the
splitting between $m_1$ and $m_2$ at low energies 
is sufficient  to generate EWSB.
The resulting SUSY mass spectrum is not very sensitive to the
two alternatives for obtaining  $m_1^2 +\Sigma_1 > m_2^2 + \Sigma_2$:
either through a splitting between $m_1$ and $m_2$ 
already at the GUT scale via D-terms
or by generating a difference  via the radiative corrections. 
%
%
%\subsection{Discussion of the remaining constraints}
\vspace{0.2cm}\\
{\bf\underline{Discussion of the remaining constraints}}\vspace{0.2cm}\\
In fig.~\ref{\figIII} the total $\chi^2$ distribution is shown as a
function of $\mze$ and $\mha$ for the two values of $\tb$ determined
above. One observes  minima at $\mze,\mha$ around (200,270) and
(800,90), as indicated by the stars. These curves were still 
produced with the data from last year. With the new coupling constants
one finds slightly different minima, as given in 
  table~\ref{t2}. In this case the minimum $\chi^2$ is not as good, since the
fit wants $\as\approx 0.125$, i.e. about 1.6$\sigma$ above the measured
LEP value and the calcaluted $\bsg$ rate is above the experimental value too,
if one takes as renormalization scale $\mu\approx 0.65 m_b$. At this
 scale  the next higher order corrections, as calculated by \cite{misiak},
are minimal.

The contours in fig.~\ref{\figIII} show the regions excluded by
different constraints used in the analysis:
\vspace{0.cm}\\
 \underline{\bf LSP Constraint:} 
The requirement that the LSP is
  neutral excludes the regions with small $m_0$ and relatively large
  $m_{1/2}$, since in this case one of the scalar staus becomes the LSP
  after mixing via the off-diagonal elements in the mass matrix. 
  The LSP constraint is especially effective at
  the high \tb region, since the off-diagonal element in the stau mass matrix 
  is proportional
  to $A_t m_0 - \mu\tan\beta$.
\vspace{0.2cm}\\
 \underline{\bf \bsg Rate:} 
  At low \tb the \bsg rate is close
  to its SM value for most of the plane. The charginos and/or the
  charged Higgses are only light enough at small values of $m_0$ and
  $m_{1/2}$ to contribute significantly. The trilinear couplings
  were found to play a negligible role for low \tb. 
  However, for large $\tb$ the trilinear coupling needs to be
  left free, since it is difficult to fit simultaneously $\bsg$, $m_b$
  and $m_\tau$. The reason is that the corrections to $m_b$ are
  large for large values of \tb due to the large contributions
  from $\tilde{g}-\tilde{q}$ and $\tilde{\chi}^\pm - \tilde{t}$
  loops proportional to $\mu\tb$. They 
  become of the order of 10-20\%. In
  order to obtain $m_b(M_Z)$ as low as 2.84 GeV, these corrections
  have to be negative, thus requiring $\mu$ to be negative.
  The \bsg rate is too large in
  most of the parameter region for large \tb, because of the
dominant chargino contribution, which is proportional to $A_t\mu$.
%\begin{eqnarray}
%\label{chargino}
%  A_{\gamma,g} & \sim &
%       \frac{m^2_t}{m^2_{\tilde{t}}}
%       \frac{A_t\mu}{m^2_{\tilde{t}}}\tan\beta.
%\end{eqnarray}
For positive (negative) values of $A_t\mu$ this leads to a larger
(smaller) branching ratio \Bbsg than  for the Standard Model with two
Higgs doublets.
In order to reduce
  this rate one needs $A_t(M_Z)>0$ for $\mu<0$.
  Since for large \tb $A_t$ does not show a fix point behaviour,
  this is possible.  
\vspace{0.2cm}\\
 \underline{\bf Relic Density:}
  The long lifetime of the universe requires a  mass density below the
  critical density, else the overclosed universe would have collapsed long ago.
  This requires that the contribution from the LSP to the relic density has to be 
  below the critical density, which can be achived if the annihilation rate
  is high enough. Annihilation into electron-positron pairs proceeds either
  through t-channel selectron exchange or through s-channel $Z^0$ exchange
  with a strength given by the Higgsino component of the lightest neutralino.
  For the low \tb scenario the value of
  $\mu$ from EWSB is large\cite{wezp}. In this case
  there is little mixing between the higgsino- and gaugino-type neutralinos as is
  apparent from the neutralino mass matrix: for $|\mu| \gg M_1 \approx
  0.4 m_{1/2}$ the mass of the LSP is simply $0.4 m_{1/2}$ and the
  ``bino'' purity is 99\%\cite{wezp}. For the high \tb
  scenario $\mu$ is much smaller  and the
  Higgsino admixture becomes larger.  This leads to an enhancement of
  $\tilde\chi^0-\tilde\chi^0$ annihilation via the s-channel Z boson
  exchange, thus reducing the relic density.  As a result, in the
  large $\tb$ case the constraint $\Omega h_0^2 < 1$ is almost always
  satisfied unlike in the case of low $\tb$.
%
%
%----------------- mh vs. mtop
%
%
\begin{figure}[t]
  \vspace*{-.35cm}
  \begin{center}
    \leavevmode
    \epsfxsize=8.5cm
    \epsffile{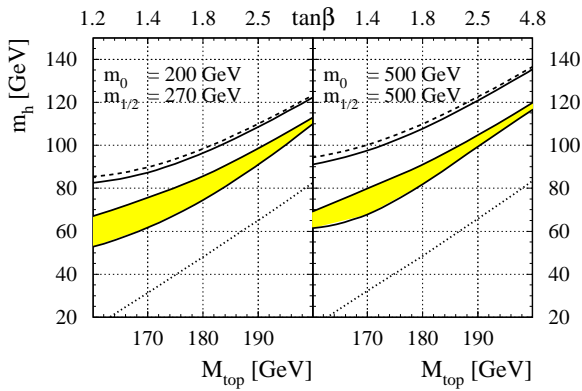}
  \end{center}
%  \vspace*{-0.7cm}
  \caption[]{\label{f4}The mass of the lightest CP-even Higgs as 
    function of the top mass at Born level (dotted lines), 
    including complete one-loop contributions of all particles
    (dashed lines). Two-loop contributions reduce the one-loop
    corrections significantly as shown by the dashed area
    (the upper boundary corresponds to $\mu>0$, the lower one
    to $\mu<0$). The solid line just
    below the dashed line is the one-loop prediction 
    from the third generation only,
    which apparently gives the main contribution. The upper scale
    indicates the value of $\tb$, as calculated from the top mass
    by the requirement of $b\tau$-unification.
    }
\end{figure}
\section{Discovery Potential  at LEP II}
Table \ref{t2} shows that charginos, neutralinos and the lightest
Higgs belong to the lightest particles in the MSSM.
Charginos are expected to  be easy to discover, since they will be pair produced
with a large cross section of several $pb$ and
lead to events with characteristic decays similar to $W^\pm$ pairs
plus missing energy.
 
The Higgs mass  depends on the top mass as shown in fig. \ref{f4}.
Here the most significant second order corrections to the Higgs mass
have been incorporated \cite{ll}, which reduces the Higgs mass by
about 15 GeV \cite{bekhiggs}. In this case 
the Higgsmass is below 90 GeV, provided the top mass is below 180 GeV (see fig. \ref{f4}),
which implies that 
the foreseen LEP energy of
192 GeV is sufficient to cover the whole parameter space.
\section{Summary}
%In summary,	
%in the Constrained Minimal Supersymmetric Model (CMSSM) the optimum
%values of the GUT scale parameters and the corresponding SUSY mass
%spectra for the low and high $\tb$ scenario have been determined from
%a combined fit to the low energy data on couplings, quark and lepton
%masses of the third generation, the electroweak scale $\mz$, $\bsg$, and
%the lifetime of the universe.
%  
The new precise determinations of the strong coupling constant $\as=0.120\pm0.003$ are slightly below the preferred CMSSM 
fit value of about 0.125. In addition,  the $\bsg$ value
of $(2.32\pm0.6)10^{-4}$ is below the predicted   values, 
at least for the SM  $(3.2\cdot10^{-4})$
and low $\tb$ scenario of the MSSM. 
For high $\tb$ the  gluino-neutralino loop
can decrease $\bsg$ somewhat.  

The lightest particles  preferred by these fits 
are charginos and higgses.
The latter has a mass below 90 GeV for a top mass below 180 GeV
in the low $\tb$ scenario, which is within reach of LEP II.

It should be noted that recent speculation about evidence for SUSY from
the $ee\gamma\gamma$ event observed by  the CDF collaboration\cite{kane}, 
the too high value of $R_b$\cite{blondel,pok,wim} and the ALEPH 4-jet events\cite{jesus,pok}
all pointed to a SUSY parameter space  inconsistent with the CMSSM, since they
require very light sparticles  (selectron, stop, chargino and/or neutralino).
However, the $R_b$ anomaly has practically disappeared\cite{blondel,wim}
and the ALEPH 4-jet events observed at 135 GeV have not   been confirmed at 161 GeV\cite{lepc}.
%\small
%\section{Acknowledgements}
%
\vspace{0.15cm}\\
{\bf\underline{Acknowledgments:}}\vspace{0.2cm}\\
Thanks  go to my close 
collaborators 
 R. Ehret, D. Kazakov, and  Ulrich Schwickerath during this analysis.
This work was  partly done during a sabbatical and
support  from the Volkswagen-Stiftung (Contract I/71681)
is greatly appreciated. 
%The research described in this publication was made possible in part
%by support from the Human Capital and Mobility Fund (Contract
%ERBCHRXCT 930345) from the European Community, and by support from the
%German Bundesministerium f\"ur Bildung und Forschung (BMBF) (Contract
%05-6KA16P) and from the Deutsche Forschungs-Gemeinschaft (DFG) for the
%Graduiertenkolleg  in Karlsruhe.
%
\vspace{0.2cm}\\
{\bf

%\addcontentsline{toc}{chapter}{References.} 
%   \bibliographystyle{unsrt}
%   \bibliography{biblio}

\underline{References:}}
\begin{thebibliography}{10}
%
\small  
\bibitem{rev}
{\em {\rm For references see e.g. the review papers: \\ H.-P. Nilles, Phys. Rep.
  {\bf 110} (1984) 1;\\ H.E. Haber, G.L. Kane, Phys. Rep. {\bf 117} (1985)
  75; 
%A.B. Lahanas and D.V. Nanopoulos, Phys. Rep. {\bf 145} (1987) 1;  R.
%  Barbieri, Riv. Nuo. Cim. {\bf 11} (1988) 1.
W.~de~Boer,
 {\em Progr. in Nucl. and Particle Phys., {\bf 33} (1994) 201}.}}
\bibitem{schmelling} M. Schmelling, these proceedings.
\bibitem{hollik1} W. de Boer, A. Dabelstein, W. Hollik, W. M\"osle and U. Schwickerath,
hep-ph/9609209
\bibitem{blondel} A. Blondel, these proceedings.
\bibitem{misiak} M. Misiak, these proceedings.
\bibitem{wezp} W. de Boer et al., Z. Phys. {\bf C71} (1996) 415.
%
\bibitem{abf1}
U.~Amaldi, W.~de~Boer, and H.~F\"urstenau, {\em Phys.~Lett.~{\bf B260} (1991) 447;}
%
\bibitem{abf2}
U.~Amaldi,et al., 
%W.~de~Boer, P.~H. Frampton, H.~F\"urstenau, and J.T. Liu,
{\em Phys.~Lett.~{\bf B281} (1992) 374.}
%{\em {\rm H.~Murayama and T.~Yanagida,} Mod. Phys. Lett. {\bf A7} (1992) 147;
%  {\rm T.~G.~Rizzo}, Phys.~Rev.~{\bf D45} (1992) 3903; {\rm T.~Moroi,
%  H.~Murayama and T.~Yanagida}, Phys. Rev. {\bf D48} (1993) 2995.}
%
%
\bibitem{pok}H. Murayama, M. Olechowski, and S. Pokorski, Phys. Lett. {\bf B371} (1996) 57
and ref. therein.
\bibitem{bekhiggs}
A.V.~Gladyshev et~al.
 {\em IEKP-KA/96-03, hep-ph/9603346 and
  references therein.}
\bibitem{ll}M.Carena et al.,
 {\em  hep-ph/9602250}
\bibitem{kane}G. Kane, these proceedings; S. Dimopoulos et al, Phys. Rev. {\bf D54} (1996) 3283.
\bibitem{wim}W. de Boer, these proceedings.
\bibitem{jesus} J. Marco, these proceedings.
\bibitem{pok} S. Pokorski, these proceedings.
\bibitem{lepc} Presentations by the 4 LEP Collaborations at the LEPC Meeting, CERN, Oct. 8, 1996. 
\end{thebibliography}
\end{document}